\documentclass[letterpaper,twocolumn,prl,aps,superscriptaddress,amsmath,amssymb,floatfix]{revtex4-1}
\usepackage{mathptmx}
\usepackage[utf8]{inputenc}
\usepackage{color}
\usepackage{amsmath}
\usepackage{amssymb}
\usepackage{graphicx}
\usepackage{esint}
\usepackage{ulem}
\usepackage[unicode=true,
 bookmarks=true,bookmarksnumbered=false,bookmarksopen=false,
 breaklinks=false,pdfborder={0 0 1},backref=false,colorlinks=true]
 {hyperref}
\hypersetup{
 linkcolor=magenta,urlcolor=blue,citecolor=blue,pdfstartview={FitH},hyperfootnotes=false}

\makeatletter

\DeclareFontEncoding{LGR}{}{}

\ProvideTextCommand{\~}{LGR}[1]{\char126#1}

\newcommand{\lyxmathsym}[1]{\ifmmode\begingroup\def\b@ld{bold}
  \text{\ifx\math@version\b@ld\bfseries\fi#1}\endgroup\else#1\fi}

\usepackage{textcomp}
\usepackage{epstopdf}

\usepackage{amsfonts}

\usepackage{soul}

\pdfpageheight\paperheight
\pdfpagewidth\paperwidth

\@ifundefined{textcolor}{}{%
 \definecolor{BLACK}{gray}{0}
 \definecolor{WHITE}{gray}{1}
 \definecolor{RED}{rgb}{1,0,0}
 \definecolor{GREEN}{rgb}{0,1,0}
 \definecolor{BLUE}{rgb}{0,0,1}
 \definecolor{CYAN}{cmyk}{1,0,0,0}
 \definecolor{MAGENTA}{cmyk}{0,1,0,0}
 \definecolor{YELLOW}{cmyk}{0,0,1,0}
}

\usepackage{xcolor}\usepackage{soul}
\newcommand{\ket}[1]{\ensuremath{\left|#1\right\rangle}}

\definecolor{blue}{rgb}{0,0,1}
\definecolor{red}{rgb}{1,0,0}
\definecolor{green}{rgb}{0,1,0}

\makeatother

\begin{document}
\affiliation{Key Laboratory of Quantum Information, Chinese Academy of Sciences,
University of Science and Technology of China, Hefei 230026, P. R. China.}
\affiliation{CAS Center For Excellence in Quantum Information and Quantum Physics,
University of Science and Technology of China, Hefei, Anhui 230026,
P. R. China.}

\title{Optimized readout strategies for neutral atom quantum processors}

\author{Liang~Chen}
\affiliation{Key Laboratory of Quantum Information, Chinese Academy of Sciences,
University of Science and Technology of China, Hefei 230026, P. R. China.}
\affiliation{CAS Center For Excellence in Quantum Information and Quantum Physics,
University of Science and Technology of China, Hefei, Anhui 230026,
P. R. China.}

\author{Wen-Yi~Zhu}
\affiliation{Key Laboratory of Quantum Information, Chinese Academy of Sciences,
University of Science and Technology of China, Hefei 230026, P. R. China.}
\affiliation{CAS Center For Excellence in Quantum Information and Quantum Physics,
University of Science and Technology of China, Hefei, Anhui 230026,
P. R. China.}

\author{Zi-Jie~Chen}
\affiliation{Key Laboratory of Quantum Information, Chinese Academy of Sciences,
University of Science and Technology of China, Hefei 230026, P. R. China.}
\affiliation{CAS Center For Excellence in Quantum Information and Quantum Physics,
University of Science and Technology of China, Hefei, Anhui 230026,
P. R. China.}

\author{Zhu-Bo~Wang}
\email{zbwang@ustc.edu.cn}
\affiliation{Key Laboratory of Quantum Information, Chinese Academy of Sciences,
University of Science and Technology of China, Hefei 230026, P. R. China.}
\affiliation{CAS Center For Excellence in Quantum Information and Quantum Physics,
University of Science and Technology of China, Hefei, Anhui 230026,
P. R. China.}

\author{Ya-Dong~Hu}
\affiliation{Key Laboratory of Quantum Information, Chinese Academy of Sciences,
University of Science and Technology of China, Hefei 230026, P. R. China.}
\affiliation{CAS Center For Excellence in Quantum Information and Quantum Physics,
University of Science and Technology of China, Hefei, Anhui 230026,
P. R. China.}

\author{Qing-Xuan~Jie}
\affiliation{Key Laboratory of Quantum Information, Chinese Academy of Sciences,
University of Science and Technology of China, Hefei 230026, P. R. China.}
\affiliation{CAS Center For Excellence in Quantum Information and Quantum Physics,
University of Science and Technology of China, Hefei, Anhui 230026,
P. R. China.}

\author{Guang-Can~Guo}
\affiliation{Key Laboratory of Quantum Information, Chinese Academy of Sciences,
University of Science and Technology of China, Hefei 230026, P. R. China.}
\affiliation{CAS Center For Excellence in Quantum Information and Quantum Physics,
University of Science and Technology of China, Hefei, Anhui 230026,
P. R. China.}
\affiliation{Hefei National Laboratory, University of Science and Technology of China, Hefei 230088, China}

\author{Chang-Ling~Zou}
\email{clzou321@ustc.edu.cn}
\affiliation{Key Laboratory of Quantum Information, Chinese Academy of Sciences,
University of Science and Technology of China, Hefei 230026, P. R. China.}
\affiliation{CAS Center For Excellence in Quantum Information and Quantum Physics,
University of Science and Technology of China, Hefei, Anhui 230026,
P. R. China.}
\affiliation{Hefei National Laboratory, University of Science and Technology of China, Hefei 230088, China}
\affiliation{Anhui Center for fundamental sciences in theoretical physics, University of Science and Technology of China, Hefei, Anhui 230026,
P. R. China.}

\date{\today}

\begin{abstract}
Neutral atom quantum processors have emerged as a promising platform for scalable quantum information processing, offering high-fidelity operations and exceptional qubit scalability. A key challenge in realizing practical applications is efficiently extracting readout outcomes while maintaining high system throughput, i.e., the rate of quantum task executions. In this work, we develop a theoretical framework to quantify the trade-off between readout fidelity and atomic retention. Moreover, we introduce a metric of quantum circuit iteration rate (qCIR) and employ normalized quantum Fisher information to characterize system overall performance. Further, by carefully balancing fidelity and retention, we demonstrate a readout strategy for optimizing information acquisition efficiency. Considering the experimentally feasible parameters for $^{87}\mathrm{Rb}$ atoms, we demonstrate that qCIRs of $197.2\,\mathrm{Hz}$ and $154.5\,\mathrm{Hz}$ are achievable using single photon detectors and cameras, respectively. These results provide practical guidance for constructing scalable and high-throughput neutral atom quantum processors for applications in sensing, simulation, and near-term algorithm implementation.

\end{abstract}
\maketitle

\section{Introduction}

\begin{figure*}
\begin{centering}
\includegraphics[width=1\linewidth]{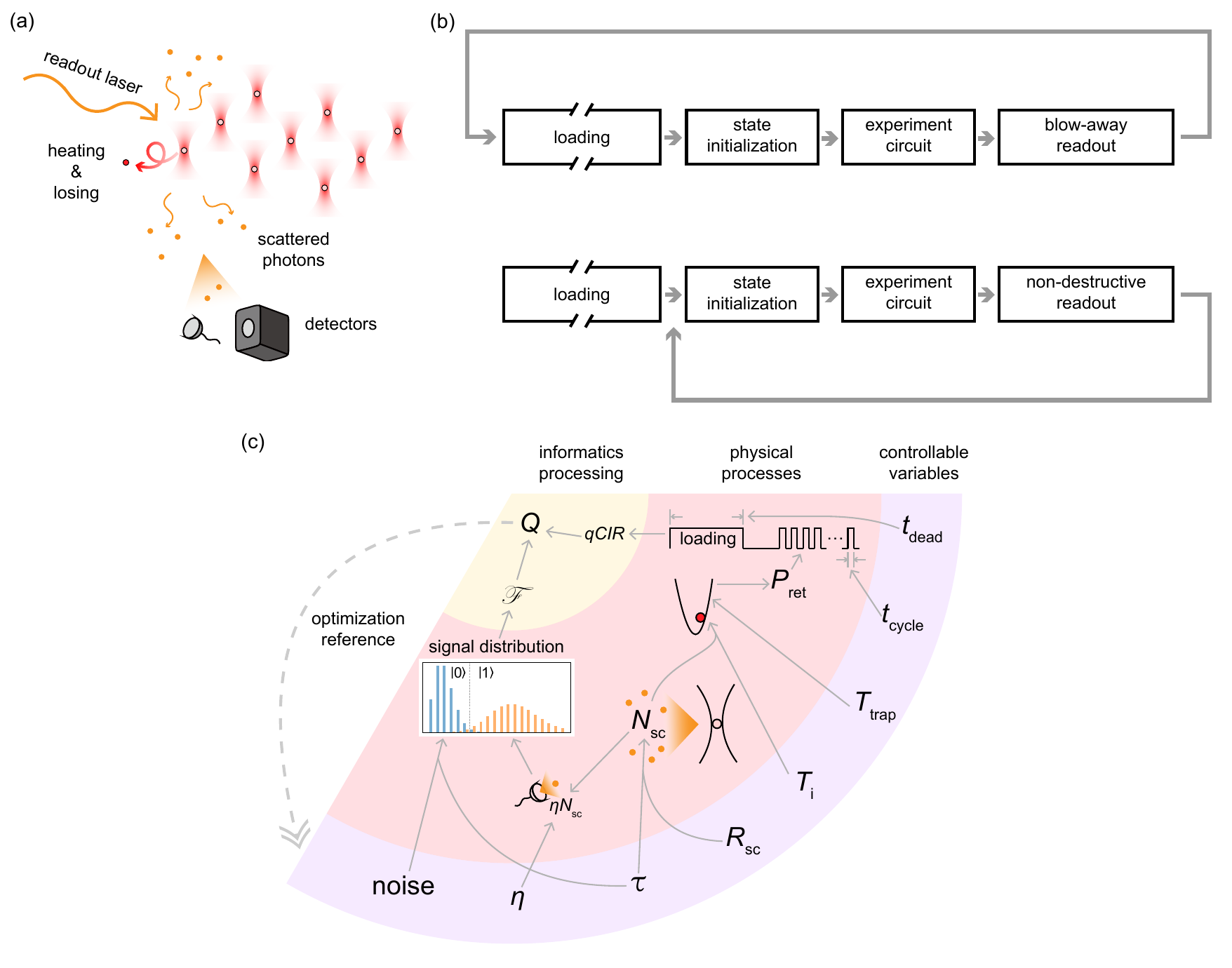}
\par\end{centering}
\caption{Conceptual overview of optimized readout strategy. (a) A schematic diagram illustrating the photon scattering process during the readout stage. As the atom scatters more photons, the detector gathers more information, thereby improving readout fidelity. However, more scattered photons also result in greater atomic heating and increased atom loss probability. (b) Experimental sequences for two different readout approaches. The upper sequence employs a blow-away approach, in which atoms need to be re-preparation after readout. In contrast, the lower sequence utilizes a non-destructive readout technique, which preserves the atoms in the trap with high retention probability. This enables immediate execution of the next circuit iteration, effectively mitigating the detrimental impact of the lengthy loading time. (c) A logic diagram revealing the effect of various experimental parameters on readout performance. Varibles in the purple domain represent controllable experimental parameters: re-preparation duration ($t_\mathrm{dead}$), quantum task cycle time ($t_\mathrm{cycle}$), trap depth ($T_\mathrm{trap}$), initial atomic temperature ($T_\mathrm{i}$), atomic scattering rate ($R_\mathrm{sc}$), readout duration ($\tau$), and overall collection efficiency ($\eta$). The red domain illustrates the physical processes, where the total scattered photons ($N_\mathrm{sc}$) and the atomic retention probability ($P_\mathrm{ret}$) emerge as critical intermediate quantities. The yellow domain quantifies performance metrics: quantum circuit iteration rate (qCIR), readout fidelity ($\mathcal{F}$) and normalized Fisher information ($Q$). The logical relationships between variables, physical processes, and performance metrics are indicated by gray arrows.}

\label{Fig1}
\end{figure*}

Neutral atom quantum systems have emerged as a leading platform for quantum information processing~\citep{saffman2016,browaeys2020,henriet2020,wu2021,morgado2021}, particularly owing to their advantage of enabling high-fidelity quantum gates~\citep{sheng2018,mcdonnell2022,evered2023,ma2023,scholl2023}, parallel gate operations and readouts~\citep{kwon2017,levine2019,muniz2024}, and exceptional scalability in qubit expansion~\citep{barredo2018,huft2022}, all of which are crucial for realizing universal quantum computing. Central to this platform is the ability to trap and manipulate individual neutral atoms using optical tweezers. These advances have enabled the realization of large-scale quantum systems with over 6,000 neutral atoms, exhibiting coherence times exceeding 12.6 seconds~\citep{manetsch2024}, surpassing the performance of solid-state quantum platforms such as superconducting qubits and quantum dots. Furthermore, neutral atom arrays can be prepared defect-free and even reconfigured adaptively with high efficiency and speed~\citep{barredo2016,endres2016,lee2017,lam2021,sheng2022,lin2024}, in stark contrast to solid-state systems, where identifying and replacing faulty components is a time-consuming process requiring extensive testing and characterization.



Despite these advantages, neutral atom systems face unique challenges compared to solid-state platforms. Solid-state qubits, once prepared, typically offer long-term stability and rapid reset after quantum operations.~\citep{huang2020,deleon2021}. In contrast, the process of high-fidelity readout in neutral atom systems inherently disrupts the atomic array, requiring complete re-preparation before the next iteration of quantum tasks, imposing a significant overhead on the execution time~\citep{saffman2016,browaeys2020,henriet2020,wu2021,morgado2021}. The disruption stems from the common technique of using resonant light to scatter photons from atoms, allowing for state determination based on scattering rate differences between qubit states. However, the scattering process imparts momentum to the atoms, causing heating and potential ejection from the optical traps~\citep{martinez-dorantes2018,bluvstein2024}. Consequently, the readout operation in neutral atom systems necessitates a complete re-preparation before the next quantum operation can be performed. This cycle of continuous array preparation and readout places stringent limitations on experimental throughput and the overall efficiency of neutral-atom-based quantum processors.



To address these challenges, there is growing interest in developing non-destructive readout techniques that mitigate the detrimental effects of photon scattering on trapped atoms. One approach is to introduce optical cavities to enhance the collection efficiency, thereby reducing the number of scattered photons required for high-fidelity readout and minimizing atom loss~\citep{bochmann2010,deist2022,gehr2010}. Non-destructive imaging techniques have also been applied to atom arrays, enabling state readout with measurement times around $10\,\mathrm{ms}$~\citep{kwon2017,martinez-dorantes2017,martinez-dorantes2018,nikolov2023,radnaev2025}. While these advances allow for the iterative reuse of atoms without requiring array re-preparation, the scalability or the overall throughput of quantum processors remains limited. Recent experimental progress has demonstmetricrated the feasibility of this approach, achieving readout fidelity and atom retention probability both exceeding 99\% for a single atom without cavity using single-photon detectors~\citep{chow2023}, which highlights the potential for fast, state-dependent, and non-destructive readout~\citep{ma2025}. However, to fully harness the potential of neutral atom platforms, a comprehensive theoretical understanding of the trade-off between readout fidelity and atomic retention probability is essential.

In this work, we systematically investigate the readout process in neutral atom quantum systems, focusing on the trade-off between optical readout fidelity and atomic retention probability. We analyze how key experimental parameters, such as scattered photon number, trap depth, collection efficiency, and experimental cycle times, influence system throughput. By employing normalized Fisher information as a key metric for performance benchmarking, our analysis reveals key insights into optimizing the neutral atom quantum processors. Specifically, we consider two readout schemes with single-photon detectors and cameras, and find that 100\,Hz-level normalized Fisher information can be achieved in both schemes, even with relatively low readout fidelity. By bridging the gap between performance metrics and system parameters, together with understanding the trade-off between fidelity and retention probability under practical experimental constraints, we provide a powerful tool for designing high-performance neutral atom quantum processors, paving the way towards demonstrating quantum advantage and addressing classically intractable practical problems.

\section{System overview}
Figure~\ref{Fig1}(a) depicts the experimental setup of a neutral atom array system~\citep{evered2023,muniz2024,sheng2018,barredo2018,huft2022,manetsch2024,barredo2018,barredo2016,huft2022,manetsch2024,sheng2022,lin2024}, highlighting the process of photon scattering during the readout stage. Atoms trapped in optical dipole traps (tweezers) are encoded into two long-coherence ground states, $\ket{0}$ and $\ket{1}$. The $\ket{1}$ state, also named the bright state, continuously scatters photons from a near-resonant readout laser, while the $\ket{0}$ state denotes the dark state, as it does not scatter the laser~\citep{bluvstein2024,gehr2010,bochmann2010,gibbons2011,fuhrmanek2011,kwon2017,martinez-dorantes2017,martinez-dorantes2018,shea2020,deist2022,chow2023,nikolov2023,radnaev2025}. This difference in scattering behavior enables state discrimination during the readout process. In the experimental sequences shown in Fig.~\ref{Fig1}(b), a typical workflow is outlined for executing quantum information processing on the neutral atom array. The procedure begins with preparing an ensemble of cold atoms via a magneto-optical trap. Then atoms are transferred into optical dipole traps, followed by a rearrangement stage to create a defect-free atom array. This loading process is necessary but time-consuming, resulting in a period of dead time during which no quantum tasks can be implemented. After the qubit array is prepared, the quantum tasks are implemented in a sequence of state initialization, quantum circuits consist of a series of quantum gates and end with a readout of qubit states. 

However, the direct discrimination between the $\ket{0}$ and $\ket{1}$ states poses a physical challenge. During photon scattering, atoms in the $\ket{1}$ state experience heating and may escape from the optical tweezers, introducing ambiguity in discriminating it from the $\ket{0}$ state. In most previous experiments, the readout is implemented destructively using the blow-away approach~\citep{bluvstein2022,graham2022,bluvstein2024}, in which the atoms in the $\ket{1}$ state are intended to be pushed out from the tweezers near-deterministically through photon scattering. The remaining atoms, assumed to be in the $\ket{0}$ state, are then detected via a photon-scattering process with a cooling effect, which can achieve high fidelity but is slow and indistinguishable in states. In both cases, the state of the atoms or the existence of the atoms is readout by collecting and detecting the scattered photons via single-photon resolved devices, among which single photon detector (SPD)~\citep{gibbons2011,shea2020,fuhrmanek2011,chow2023} and qCMOS camera~\citep{bluvstein2024,manetsch2024} have become common and promising options.

Instead of perusing the high readout fidelity for single-shot measurement, we prefer the non-destructive readout of atomic states~\citep{bochmann2010,gehr2010,gibbons2011,fuhrmanek2011,kwon2017,martinez-dorantes2017,martinez-dorantes2018,shea2020,deist2022,chow2023,nikolov2023,radnaev2025}. As shown in Fig.~\ref{Fig1}(b), the blow-away readout requires resetting the entire tweezer array for another cycle of quantum tasks. In contrast, the non-destructive readout allows the reuse of the tweezer array without the time-consuming tweezer array preparation process. Due to the photon recoil effect, there is no perfect non-destructive readout, and the trade-off between obtaining high-fidelity information and the detrimental effects of heating must be carefully managed.

Figure~\ref{Fig1}(c) illustrates the detailed physical processes involved in the readout and the overall operation of neutral atom quantum systems including various parameters that influence the system's performance: 
\begin{enumerate}

    \item Heating of the atoms in the trap ($\tau$, $R_{\mathrm{sc}}$, $T_{\mathrm{i}}$, $T_{\mathrm{trap}}$). The photon scattering process causes heating of the trapped atom. The increase in temperature for each absorption or emission of single photons due to the recoil effect~\citep{martinez-dorantes2018} is $\Delta T_\mathrm{ph}={p_\mathrm{ph}^2}/{2k_\mathrm{B}m_\mathrm{a}}$, where $k_\mathrm{B}$ is the Boltzmann constant, $p_\mathrm{ph}$ is the momentum of a readout laser photon, $m_\mathrm{atom}$ is the mass of the atom.  With a rate $R_\mathrm{sc}$ and a duration of $\tau$, i.e., a total number of $N_\mathrm{sc}=R_{\mathrm{sc}}\tau$ scattered photons, the atom temperature is increased from $T_\mathrm{i}$ to  $T_\mathrm{a}=T_\mathrm{i}+2N_\mathrm{sc} \Delta T_\mathrm{ph}$, and the probability of atom retention in the trap is determined by both $T_\mathrm{a}$ and the trap depth $T_{\mathrm{trap}}$.
    \item State discrimination ($\eta$, $\tau$, $R_{\mathrm{sc}}$, detector noise). The discrimination of the internal state of the atoms is determined by the collected photons. On the one hand, the atom in $\ket{1}$ state scatters $N_{\mathrm{sc}}$ photons from the readout laser, while only part of it can be detected ($\eta$) due to limited lens N.A., optical path transmittance and detector quantum efficiency. On the other hand, the detector has noise counts even if the atoms are in the $\ket{0}$ state, mainly due to the electronic dark counts of the detector. 
    \item Repetition of the experimental sequences ($t_{\mathrm{dead}}$, $t_{\mathrm{cycle}}$). As illustrated in Fig.~\ref{Fig1}(b), the loading process corresponds to a dead time ($t_\mathrm{dead}$) of the quantum system. $t_{\mathrm{cycle}}$ is the time for implementing one cycle of a quantum task. 
\end{enumerate}

These parameters play a crucial role in the overall performance of the neutral atom quantum computing system. The optimization requires a detailed understanding of the physical processes involved in each stage, particularly the trade-off between readout fidelity and atomic heating. Moreover, a credible evaluation of the system's efficiency in extracting information is necessary.

\begin{figure*}
\begin{centering}
\includegraphics[width=1\linewidth]{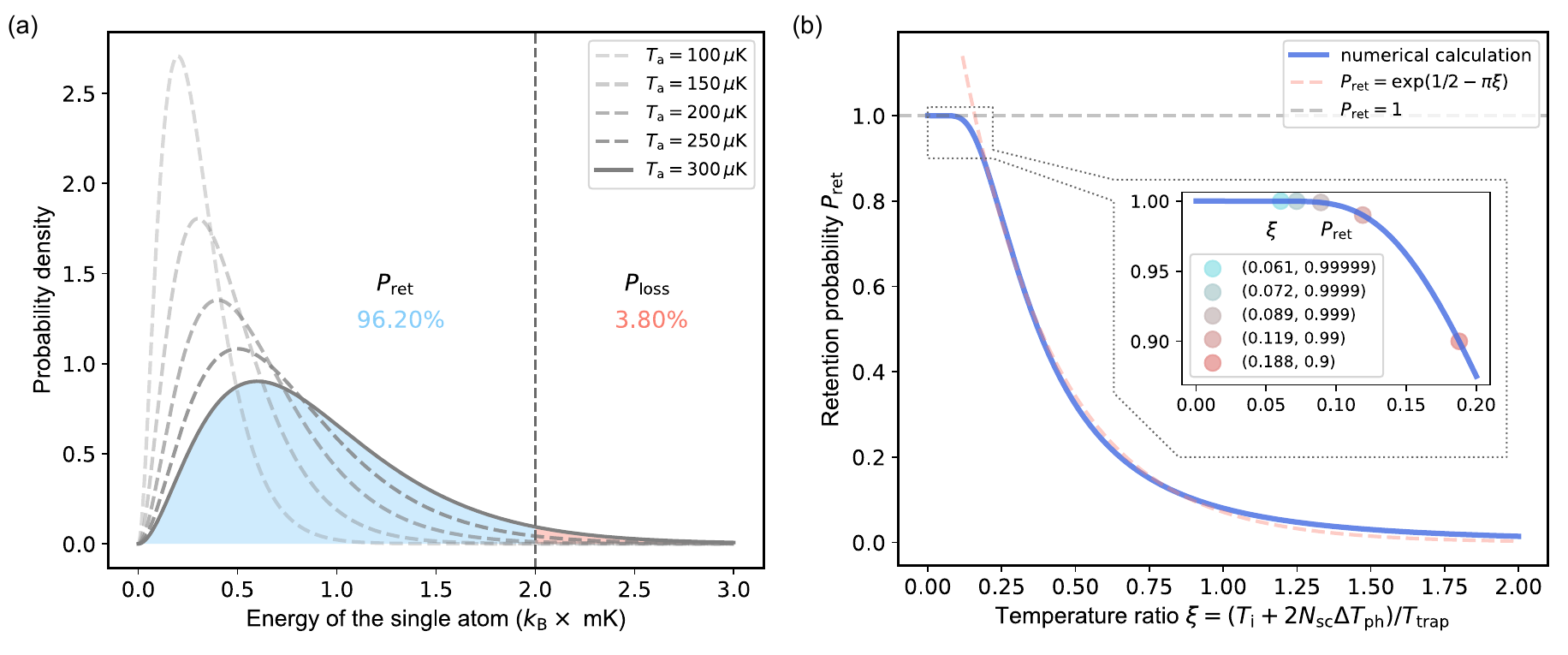}
\par\end{centering}
\caption{Trade-off between $N_\mathrm{sc}$ and $P_\mathrm{ret}$. (a) Maxwell-Boltzmann energy distribution for an atom at temperatures of $T_\mathrm{a}=100,\,150,\,200,\,250,\,300\,\mu\mathrm{K}$. The retention probability $P_\mathrm{ret}$ (loss probability $P_\mathrm{loss}$) is derived from the integral of the distribution below (above) the trap depth $T_\mathrm{trap}$, which is hinted by the blue (red) shaded area. Numerical values are calculated for an atom temperature of $300\,\mu\mathrm{K}$ and a trap depth of $2\,\mathrm{mK}$. (b) Dependence of retention probability $P_\mathrm{ret}=1-P_\mathrm{loss}$ on the temperature ratio $\xi=T_\mathrm{a}/T_\mathrm{trap}$. The inset shows several specific points for reference. }
\label{Fig2}
\end{figure*}

\section{Model}
\subsection{Heating and atom loss}
For a single trapped atom with a temperature of $T_\mathrm{a}$, its mechanical energy follows the Maxwell-Boltzmann distribution~\cite{tuchendler2008} {[}see Fig.$\,$\ref{Fig2}(a){]}:
\begin{equation}
f_\mathrm{temp}(t,T_{\mathrm{a}})=\frac{t^2}{2T_\mathrm{a}^3}e^{-t/T_\mathrm{a}}.\label{eq:1}
\end{equation}
The probability of this atom escaping from a trap with depth $T_\mathrm{trap}$ is calculated by the integral of this distribution above the trap depth $T_\mathrm{trap}$, i.e.
\begin{align}
P_\mathrm{loss}=\int_{T_\mathrm{trap}}^{+\infty} f_\mathrm{temp}(t,T_{\mathrm{a}})dt=\left(1+\frac{1}{\xi}+\frac{1}{2\xi^2}\right)e^{-\frac{1}{\xi}},\label{eq:2}
\end{align}
where  $\xi=T_\mathrm{a}/T_\mathrm{trap}$ is the ratio between the atom temperature $T_\mathrm{a}=T_\mathrm{i}+2N_\mathrm{sc} \Delta T_\mathrm{ph}$ and the trap depth. As shown by the shaded area in Fig.~\ref{Fig2}(a), the atom retention probability $P_\mathrm{ret}=1-P_\mathrm{loss}=96.2\%$ for an atom temperature of $T_\mathrm{a}=300\,\mu\mathrm{K}$ and a trap depth of $T_\mathrm{trap}=2\,\mathrm{mK}$. Figure~\ref{Fig2}(b) shows the dependence of the atom retention probability $P_\mathrm{ret}=1-P_\mathrm{loss}$ on the temperature ratio. We found that for $\xi<0.1$, the loss is negligible as $P_\mathrm{loss}\approx \frac{1}{2\xi^2}e^{-\frac{1}{\xi}}$, which decays exponentially fast to zero, with the dominant factor being $e^{-\frac{1}{\xi}}$. The inset in Fig.$\,$\ref{Fig2}(b) shows a zoomed-in view around this plateau and gives specific values for several points of interest, with $P_\mathrm{loss}$ suppressed by one order of magnitude when simply reducing the temperature ratio by $20\%$. For larger $\xi$, the $P_\mathrm{ret}$ decays very fast, exhibiting an approximate exponentially decreasing behavior. 

The core physics of our readout strategy is a moderate selection of the experimental parameters to keep the atom temperature within the plateau after each readout stage, i.e. to minimize $\xi$. This set a corresponding limit on the allowed number of scattered photons as 
\begin{equation}
N_\mathrm{sc}=\frac{\xi T_\mathrm{trap}-T_\mathrm{i}}{2\Delta T_\mathrm{ph}},\label{eq:3}
\end{equation}
which imposes the accessible information for state readout.

\subsection{Readout fidelity}
\begin{figure}
\begin{centering}
\includegraphics[width=1\linewidth]{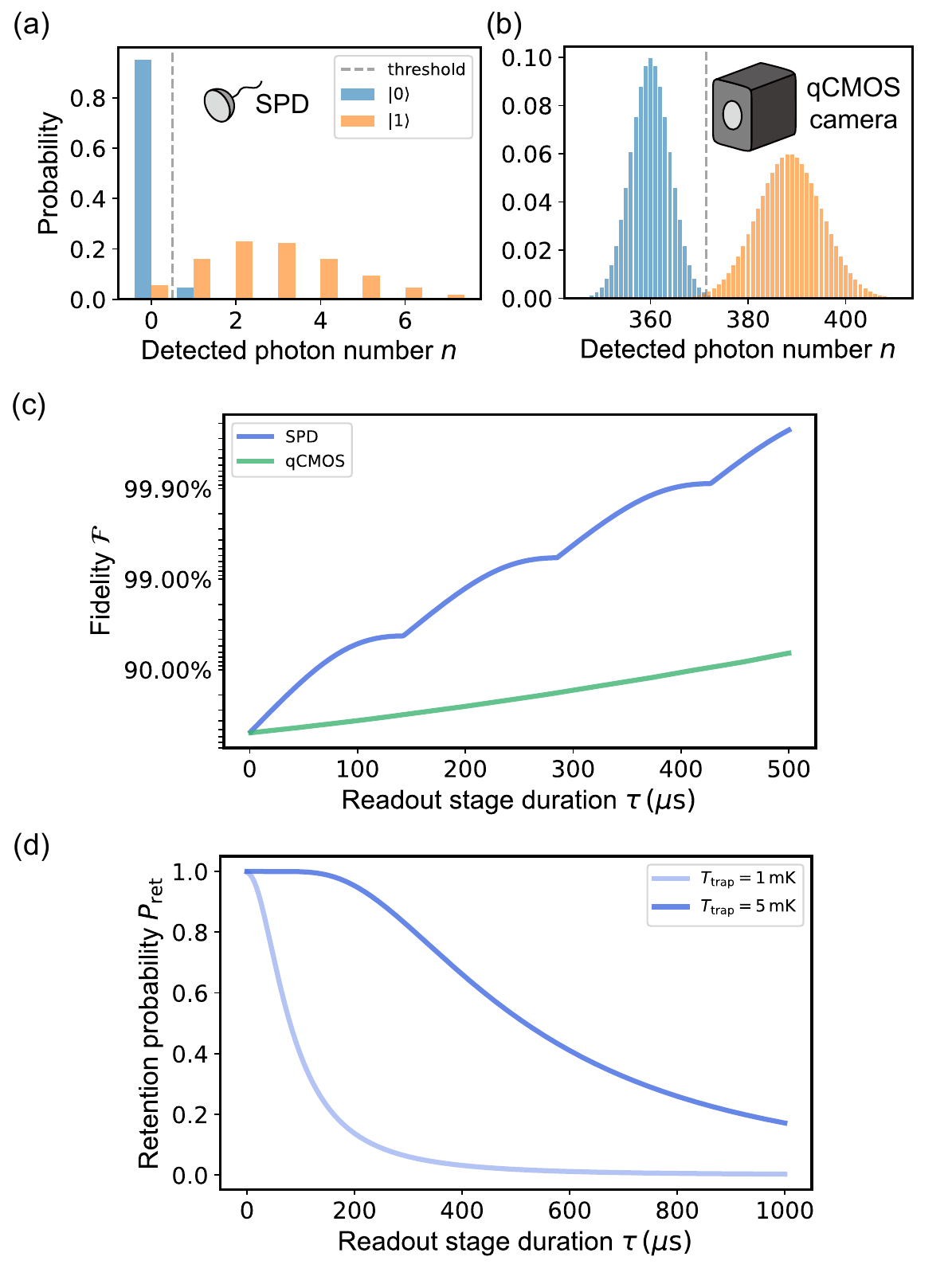}
\par\end{centering}
\caption{Calculation results of $\mathcal{F}$ and $P_\mathrm{ret}$. (a) Photon-counting statistics for SPD readout with dark count $N_\mathrm{D,SPD}=0.05$ and atom fluorescence signal $\eta N_\mathrm{sc}=2.85$. (b) Photon-counting statistics for qCMOS camera readout with readout noise $N_\mathrm{QC}=360$, $\sigma_\mathrm{QC}=4$ and atom fluorescence signal $\eta N_\mathrm{sc}=28.5$. (c) Fidelity $\mathcal{F}$ versus readout stage duration $\tau$ for SPD and qCMOS. Curves are calculated with related parameters set to $(R_\mathrm{D,SPD},\,N_\mathrm{QC},\,\sigma_\mathrm{QC},\,\eta,\,R_\mathrm{sc})=(500\,\mathrm{Hz},\,360,\,4,\,0.3\,\%,9.5\,\mathrm{MHz})$. (d) Retention probability $P_\mathrm{ret}$ as a function of $\tau$ for shallow ($T_\mathrm{trap}=1\,\mathrm{mK}$) and deep ($T_\mathrm{trap}=5\,\mathrm{mK}$) optical traps with $T_\mathrm{i}=100\,\mu\mathrm{K}$ and $R_\mathrm{sc}=9.5\,\mathrm{MHz}$.}
\label{Fig3}
\end{figure}

The two types of detectors show similar detection efficiency around $50\%$, while their background noises are essentially different. For SPD, its noise is mainly from the dark counts of this optoelectronic device, which appears as a Poisson distribution with probability
\begin{align}
g_{\ket{0},\mathrm{SPD}}(n)=\frac{e^{-R_\mathrm{D,SPD}\tau}(R_\mathrm{D,SPD}\tau)^n}{n!},\label{eq:4}
\end{align}
for $n$ clicks within a detection duration of $\tau$ with dark count rate $R_\mathrm{D,SPD}$. Similarly, for the $\ket{1}$ state, the counts follow the Poisson distribution with a higher mean photon number as 
\begin{equation}
g_\mathrm{\ket{1},SPD}(n)=\frac{e^{-(R_\mathrm{D,SPD}+\eta R_\mathrm{SC})\tau}[(R_\mathrm{D,SPD}+\eta R_\mathrm{SC})\tau]^n}{n!}.\label{eq:5}
\end{equation}

For qCMOS camera, its noise is mainly the camera readout noise, which has a feature of high mean, relatively low variance, and time independence. The signal could be described by Gaussian probability distribution
\begin{align}
g_\mathrm{\ket{0},QC}(n)=\frac{1}{\sqrt{2\pi}\sigma_\mathrm{QC}} e^{-\frac{(n-N_\mathrm{QC})^2}{2\sigma_\mathrm{QC}^2}},\label{eq:6}
\end{align}
with $N_\mathrm{QC}$ and $\sigma_\mathrm{QC}^{2}$ denoting the mean and variance of qCMOS noise, respectively. For the $\ket{1}$ state, scatter photons with Poisson distribution added to the signal of the $\ket{0}$ state, leading to approximate Gaussian distribution {[}see Appendix A{]}
\begin{align}
g_\mathrm{\ket{1},Q}(n)=\frac{1}{\sqrt{2\pi(\sigma_\mathrm{QC}^2+\eta R_{\mathrm{SC}}\tau)}} e^{-\frac{(n-N_\mathrm{QC}-\eta R_{\mathrm{SC}}\tau)^2}{2(\sigma_\mathrm{QC}^2+\eta R_{\mathrm{SC}}\tau)}} .\label{eq:7}
\end{align}

Figures~\ref{Fig3}(a) and (b) illustrate the typical distributions of photon counts by SPD and qCMOS camera, respectively. We distinguish between the two qubit states by estimating whether the readout signal obtained is above or below a chosen threshold $n_{\mathrm{th}}$. Hence, the probability to correctly detect $\ket{1}$ state reads $P_{\ket{1},\mathrm{SPD(QC)}}=\sum_{n=n_{\mathrm{th}}}^{\infty}g_\mathrm{\ket{1},SPD(QC)}(n)$ and detect $\ket{0}$ state reads $P_{\ket{0},\mathrm{SPD(QC)}}=\sum_{n=0}^{n_{\mathrm{th}}-1}g_\mathrm{\ket{0},SPD(QC)}(n)$. 
To simplify subsequent calculations, we introduce a unified definition of readout fidelity 
\begin{equation}
    \mathcal{F}=\frac{1}{2}(P_{\ket{0},\mathrm{SPD(QC)}}+P_{\ket{1},\mathrm{SPD(QC)}}).
\end{equation}
Here, optimal $n_{\mathrm{th}}$ for SPD and qCMOS are respectively chosen to maximize the corresponding fidelities. 

To investigate the dependence of $\mathcal{F}$ on $\tau$, we assume a set of common readout device parameters of $(R_\mathrm{D,SPD},\,N_\mathrm{QC},\,\sigma_\mathrm{QC})=(500\,\mathrm{Hz},\,360,\,4)$, the rate of photon scattering $R_\mathrm{sc}=9.5\,\mathrm{MHz}$, and the total collection and detection efficiency of scattered photons $\eta=0.3\%$ for both type of detectors. The dependence of $\mathcal{F}$ on $\tau$ is shown in Fig.$\,$\ref{Fig3}(c). It can be seen that the performance of SPD is way better than that of qCMOS due to the noise difference. It is noteworthy that the curve for SPD {[}blue curve in Fig.$\,$\ref{Fig3}(c){]} exhibits distinct segmentation, which is due to the jump of optimal $n_{\mathrm{th}}$ as integers. The qCMOS curve also exhibits this phenomenon, but since the signal distribution for qCMOS is relatively broad and low, the segmentation caused by threshold jumps is less pronounced. In contrast, the signal distribution for SPD (especially the $\ket{0}$ state signal distribution when $\tau$ is small) is narrow and tall, making the segmentation caused by threshold jumps more apparent.

\subsection{Quantum circuit iteration rate}
\begin{figure}
\begin{centering}
\includegraphics[width=1\linewidth]{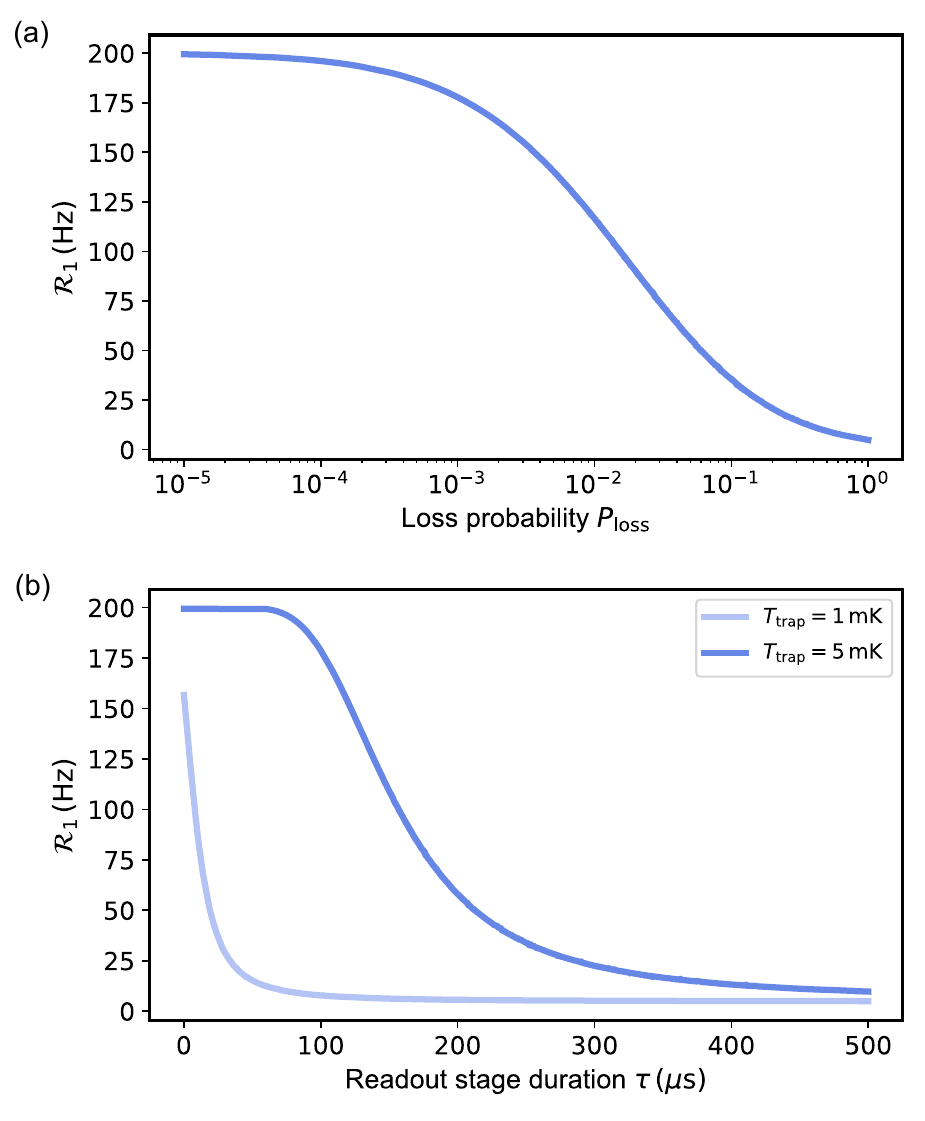}
\par\end{centering}
\caption{Calculation results of the qCIR for a single atom by the adaptive strategy ($\mathcal{R}_1$). (a) $\mathcal{R}_1$ versus loss probability $P_\mathrm{loss}$. (b) $\mathcal{R}_1$ versus readout stage duration $\tau$ with two different trap depth of $T_\mathrm{trap}=1,\,5\,\mathrm{mK}$. Other related parameters are set to $(t_\mathrm{dead},\,t_\mathrm{cycle},\,T_\mathrm{i},\,R_\mathrm{sc})=(200\,\mathrm{ms},\,5\,\mathrm{ms},\,100\,\mu\mathrm{K},\,9.5\,\mathrm{MHz})$.}
\label{Fig4}
\end{figure}
\begin{figure*}
\begin{centering}
\includegraphics[width=1\linewidth]{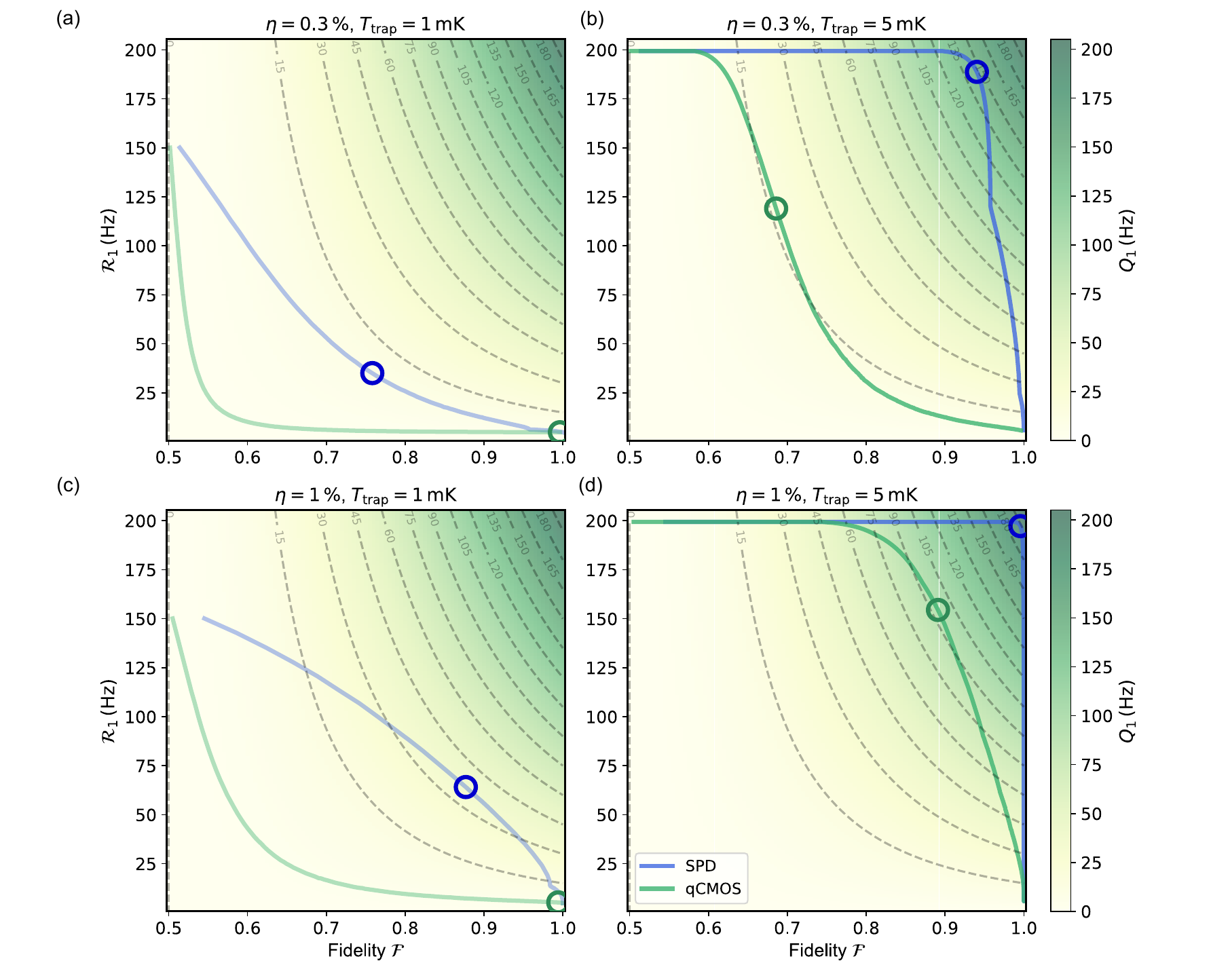}
\par\end{centering}
\caption{The trade-off characteristics between $\mathcal{R}_1$ and $\mathcal{F}$ for SPD and qCMOS camera. The related readout device parameters are set to $(R_\mathrm{D,SPD},\,N_\mathrm{QC},\,\sigma_\mathrm{QC})=(500\,\mathrm{Hz},\,360,\,4)$, and related experimental parameters are set to $(t_\mathrm{dead},\,t_\mathrm{cycle},\,T_\mathrm{i},\,R_\mathrm{sc})=(200\,\mathrm{ms},\,5\,\mathrm{ms},\,100\,\mu\mathrm{K},\,9.5\,\mathrm{MHz})$. A colormap is employed to represent $Q$ under different $\mathcal{F}$ and $\mathcal{R}_1$, with contour lines plotted for reference. The optimal $Q_1$ are explicitly marked by hollow circles. (a) $(\eta,\,T_\mathrm{trap})=(0.3\,\%,\,1\,\mathrm{mK})$. For SPD, the optimal $Q_1=9.4\,\mathrm{Hz}$ with corresponding $\mathcal{F}=75.83\,\%$ and $\mathcal{R}_1=35.1\,\mathrm{Hz}$. While for qCMOS, the optimal $Q_1=4.8\,\mathrm{Hz}$ with corresponding $\mathcal{F}=99.63\,\%$ and $\mathcal{R}_1=4.9\,\mathrm{Hz}$. (b) $(\eta,\,T_\mathrm{trap})=(0.3\,\%,\,5\,\mathrm{mK})$. For SPD, the optimal $Q_1=146.4\,\mathrm{Hz}$ with corresponding $\mathcal{F}=94.04\,\%$ and $\mathcal{R}_1=188.8\,\mathrm{Hz}$. While for qCMOS, the optimal $Q_1=16.4\,\mathrm{Hz}$ with corresponding $\mathcal{F}=68.57\,\%$ and $\mathcal{R}_1=119.1\,\mathrm{Hz}$. (c) $(\eta,\,T_\mathrm{trap})=(1\,\%,\,1\,\mathrm{mK})$. For SPD, the optimal $Q_1=36.4\,\mathrm{Hz}$ with corresponding $\mathcal{F}=87.69\,\%$ and $\mathcal{R}_1=64.1\,\mathrm{Hz}$. While for qCMOS, the optimal $Q_1=5.10\,\mathrm{Hz}$ with corresponding $\mathcal{F}=99.42\,\%$ and $\mathcal{R}_1=5.2\,\mathrm{Hz}$. (d) $(\eta,\,T_\mathrm{trap})=(1\,\%,\,5\,\mathrm{mK})$. For SPD, the optimal $Q_1=193.8\,\mathrm{Hz}$ with corresponding $\mathcal{F}=99.56\,\%$ and $\mathcal{R}_1=197.2\,\mathrm{Hz}$. While for qCMOS, the optimal $Q_1=94.5\,\mathrm{Hz}$ with corresponding $\mathcal{F}=89.11\,\%$ and $\mathcal{R}_1=154.5\,\mathrm{Hz}$.}
\label{Fig5}
\end{figure*}

Although the longer readout time could enhance the fidelity, the heating of the atom could introduce significant loss of the atom. Figure~\ref{Fig3}(d) shows the retention probability of the atoms after a period of $\tau$ readout, according to the model presented in Eq.~(\ref{eq:2}). However, $P_{\mathrm{loss}}$ can not directly reflect its impact on system performance. To gain better insights, we need to convert $P_{\mathrm{loss}}$ into the quantum circuit iteration rate (qCIR), defined by the average number of quantum circuit executions per unit of time. In the following text, we use the mark $\mathcal{R} $to represent it.

As shown in Fig.~\ref{Fig1}(b), for a relatively high $P_{\mathrm{loss}}>50\,\%$, it is necessary to re-prepare the atom array after the readout stage, which requires a time typically longer than $0.1\,\mathrm{s}$, and thus imposes an ultimate limit of qCIR to $\mathcal{R}<10\,\mathrm{Hz}$. Yet, for a low $P_{\mathrm{loss}}$, or in other words, in non-destructive readout scenarios, the atom tends to remain in the trap after the readout stage. Therefore, we can initialize its external and internal states with only a cooling stage and a pumping stage, which typically require a duration of about several $\mathrm{ms}$. Then the atom can be immediately reused for the next circuit execution, thus avoiding the overhead introduced by the atom array preparation.

For a general description of the atomic quantum system implementing quantum circuits, we have the necessary dead time, i.e., the duration to prepare the single atom array, $t_\mathrm{dead}$. The quantum circuits are implemented with the cycles, consisting of the initialization of atoms, consisting of rapid polarization gradient cooling (PGC) to reduce the atom temperature to $T_i$ again and optical pumping to initialize the qubit states, the quantum gate sequences, and the non-destructive quantum state readout, and a rapid check for existence of single atoms in tweezers, with a total duration of $t_\mathrm{cycle}\ll t_\mathrm{dead}$. Assuming that we implement $n$ iterations of quantum circuits for each preparation of the tweezer array, and the corresponding qCIR reads
\begin{align}
\mathcal{R}=\frac{n}{t_\mathrm{dead}+nt_\mathrm{cycle}}.\label{eq:9}
\end{align}
For conventional neutral atom array systems with blow-away readout, $n=1$, $\mathcal{R}\approx 1/t_\mathrm{dead}$. For an ideal non-destructive readout $P_\mathrm{loss}=0$, we have $n\gg1$, and $\mathcal{R}$ saturates to $1/t_\mathrm{cycle}$, indicating a significantly improvement of qCIR. 

For a concrete calculation of qCIR, we introduce practical experimental parameters of alkali atoms, with $t_\mathrm{dead}=200\,\mathrm{ms}$ and $t_\mathrm{cycle}=5\,\mathrm{ms}$. Here, the initialization of atoms can be implemented with a duration of $1\,\mathrm{ms}$, consisting of a rapid PGC to reduce the atom temperature to $T_i$ again and optical pumping to initialize the qubit states. Consider typical atomic coherence time at order of $\mathrm{ms}$, the gate sequences should be less than $1\,\mathrm{ms}$. Also, the non-destructive state-dependent readout can be performed within $1\,\mathrm{ms}$, and the existence check of the single atoms in the tweezer by cooling lasers can also be accomplished within $1\,\mathrm{ms}$. There are two strategies to repeat the cycles: adaptive and non-adaptive. In the first approach, we can decide whether to discard all atoms and reset the tweezer array according to the results of the existence check. For the second strategy, we can fix the number of cycle repetitions for each reset of the tweezer array, and only keep the outcomes of the quantum circuits when the existence check results are positive. We calculated the qCIR for a single atom by the adaptive strategy ($\mathcal{R}_1$) and the results are summarized in Fig.~\ref{Fig4}.  For a given atom loss rate, qCIR can be derived as 
\begin{align}
\mathcal{R}_1=\sum\limits_{n=1}^{\infty} \frac{n}{t_\mathrm{dead}+nt_\mathrm{cycle}}P_\mathrm{loss}(1-P_{\mathrm{loss}})^{n-1}.\label{eq:10}
\end{align}
Figure~\ref{Fig4}(a) shows that as $P_\mathrm{loss}$ increases, $\mathcal{R}_1$ reduces monotunely from $1/t_\mathrm{cycle}=200\,\mathrm{Hz}$ and saturated to $1/(t_\mathrm{dead}+t_\mathrm{cycle})=4.9\,\mathrm{Hz}$. Considering specific tweezer trap depths $T_\mathrm{trap}=1\,\mathrm{mK},\,5\,\mathrm{mK}$, with initial atom temperature $T_\mathrm{i}=100\,\mu\mathrm{K}$ and photon scattering rate during readout stage $R_\mathrm{sc}=0.5\times\Gamma/2\approx9.5\,\mathrm{MHz}$, the corresponding $P_{\mathrm{loss}}$ can be derived and the resulting qCIR are presented in Fig.~\ref{Fig4}(b). As expected, the $\mathcal{R}_1$ and $\mathcal{F}$ show a trade-off relation when varying the $\tau$, appealing for the quantity of the experimental performance.

\section{Normalized quantum Fisher information}

In quantum metrology and simulation applications, the quantum circuits should be repeated many times to obtain the statistics of the outcomes~\citep{madjarov2019,norcia2019,young2020,bernien2017,ebadi2021,scholl2021,semeghini2021}. The total information amount provided by the result samples decides the confidence of the output. The significance of increasing qCIR lies in obtaining sampling results at a faster rate, while improving fidelity ensures that each sampling result carries more concrete information. Therefore, to balance the trade-off between these two metrics, we quantify the information gain rate by introducing a comprehensive metric of normalized quantum Fisher information (QFI)~\citep{liu2020,meyer2021,wang2022}
\begin{align}
    Q=\mathcal{R}I,\label{eq:11}
\end{align}
with $I$ is QFI for a single cycle of the quantum circuit. By designing an appropriate task scenario and considering imperfect readout conditions {[}see Appendix B{]}, $I$ can be described by the following expression:
\begin{align}
I=(2\mathcal{F}-1)^2.\label{eq:12}
\end{align}
Therefore, we set
\begin{align}
  Q_1=\mathcal{R}_1(2\mathcal{F}-1)^2\label{eq:13}
\end{align}
as an integrated reference metric to guide the optimization of experimental system parameters for achieving the highest information acquisition rate.


In Fig.~\ref{Fig5}, the background color indicates $Q_1$ corresponding to different $\mathcal{R}_1$ and $\mathcal{F}$, and contour lines are also plotted for reference. By choosing the parameters as Fig.~\ref{Fig4}, the ultimate limit of the achievable $Q_1=200\,\mathrm{Hz}$. In practice, the corresponding achievable $\mathcal{R}_1$ and $\mathcal{F}$ are correlated, both determined by the readout time $\tau$. First, considering a relatively low atom fluorescence collection efficiency $\eta=0.3\%$ for a lens with an N.A. of around 0.3 and a trap depth of $1\,\mathrm{mK}$, the curves in Fig.$\,$\ref{Fig5}(a) show the results for SPD and qCMOS camera, respectively. Due to the trade-off between $\mathcal{R}_1$ and $\mathcal{F}$, the qCIR reduces while $\mathcal{F}$ increases with growing $\tau$. For the SPD case, although $\mathcal{F}\approx1$ is achievable by sacrificing qCIR to Hz-level, the optimal $Q_1=9.4\,\mathrm{Hz}$ is obtained with a low $\mathcal{F}=75.83\,\%$ and a relatively high $\mathcal{R}_1=35.1\,\mathrm{Hz}$. For qCMOS, due to a longer readout duration for achieving the same fidelity compared to SPD, its qCIR is lower, as shown by the green line, and the optimal $Q_1=4.8\,\mathrm{Hz}$ is obtained with a high $\mathcal{F}=99.63\,\%$ and a relatively low $\mathcal{R}_1=4.9\,\mathrm{Hz}$. Comparing the two cases, extremely different readout strategies lead to a similar normalized QFI. If we can improve the trap depth to $5\,\mathrm{mK}$, then a much lower loss of atoms can be achieved with the same amount of scattered photons, as shown in Fig.~\ref{Fig5}(b). Therefore, both SPD and qCMOS have a qCIR saturated to $200\,\mathrm{Hz}$ for a low $\mathcal{F}$, with $Q_1$ increasing with the $\mathcal{F}$. However, the performance for qCMOS is somehow insensitive to the strategy when $\mathcal{F}>65\,\%$, while the performance of SPD readout drops quickly when $\mathcal{F}$ exceeds an optimal working point ($94.04\,\%$). Comparing the case with a lower depth trap, the optimal $Q_1$ are improved to $146.4\,\mathrm{Hz}$ and $16.4\,\mathrm{Hz}$ for SPD and QCOMS, respectively. 

Further improvement is possible by enhancing the collection efficiency of the atomic fluorescences. For example, $\eta=1\%$ is achievable with a lens N.A. of around 0.5, and the corresponding results are summarized in Figs.~\ref{Fig5}(c) and (d). From the analysis of heating above, both heating and signal are determined by the collected photons $N_\mathrm{sc}$. In a rough qualitative analysis, we can consider that $T_\mathrm{i}$ is negligible compared to $T_\mathrm{trap}$. In this case, the trade-off between fidelity and qCIR is approximately determined by the combined factor $\eta T_{\mathrm{trap}}$. Thus, the combination of the parameters $(\eta,\,T_{\mathrm{trap}})=(1\%,\,1\,\mathrm{mK})$ in Fig.~\ref{Fig5}(c) outperforms $(0.3\%,\,1\,\mathrm{mK})$ in (a) and worse than $(1\%,\,1\,\mathrm{mK})$ in (b). For the parameters $(1\%,\,5\,\mathrm{mK})$ in Fig.~\ref{Fig5}(d), $Q_1$ are further improved to $193.8\,\mathrm{Hz}$ and $94.5\,\mathrm{Hz}$ for SPD and QCOMS, respectively.

\begin{figure}[t]
\begin{centering}
\includegraphics[width=1\linewidth]{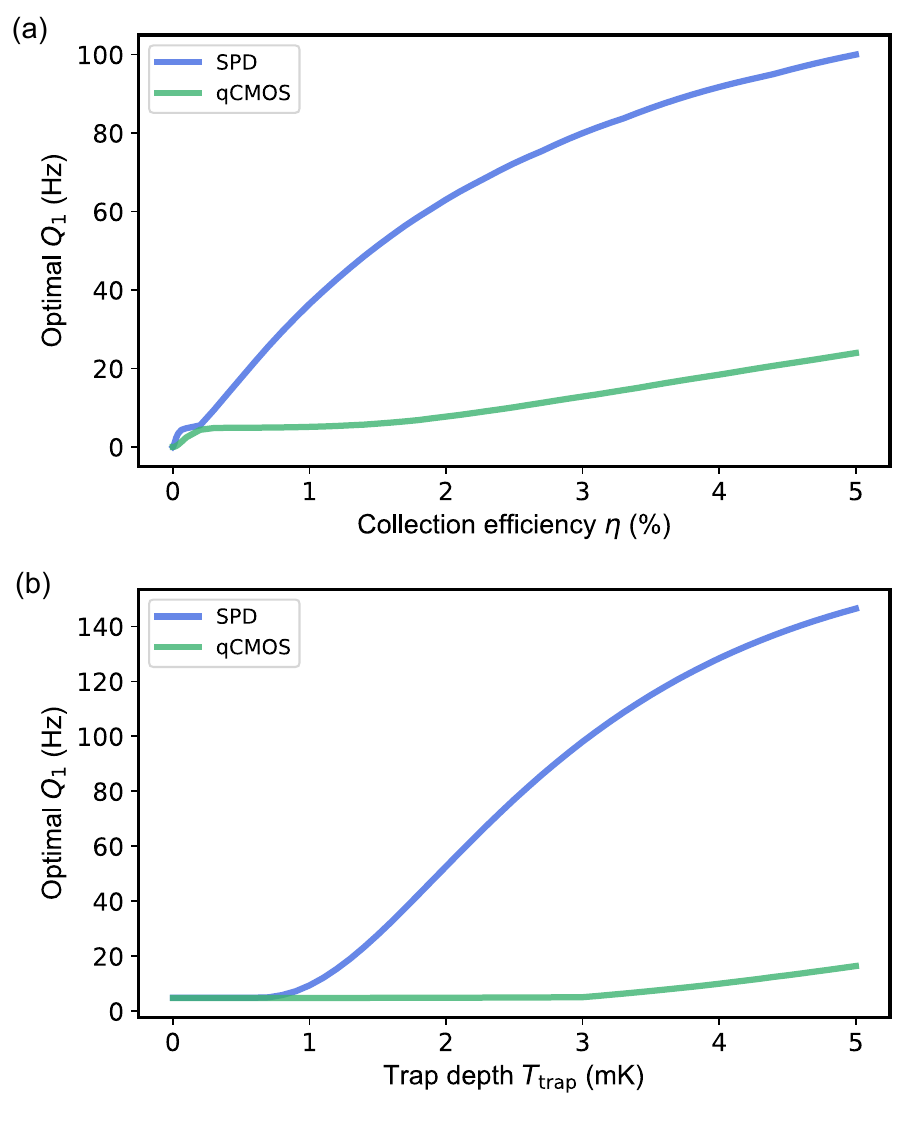}
\par\end{centering}
\caption{Investigation of optimal $Q_1$ versus collection efficiency $\eta$ and trap depth $T_\mathrm{trap}$. (a) Optimal normalized quantum Fisher information ($Q_1$) versus collection efficiency ($\eta$) with trap depth ($T_\mathrm{trap}$) set to $1\,\mathrm{mK}$. (b) Optimal normalized quantum Fisher information ($Q_1$) versus trap depth ($T_\mathrm{trap}$) with collection efficiency $\eta=0.3\,\%$. The related readout device parameters and experimental parameters in these two subfigures are the same as those in Fig.$\,$\ref{Fig5}.}
\label{Fig6}
\end{figure}

\begin{figure*}[t]
\begin{centering}
\includegraphics[width=1\linewidth]{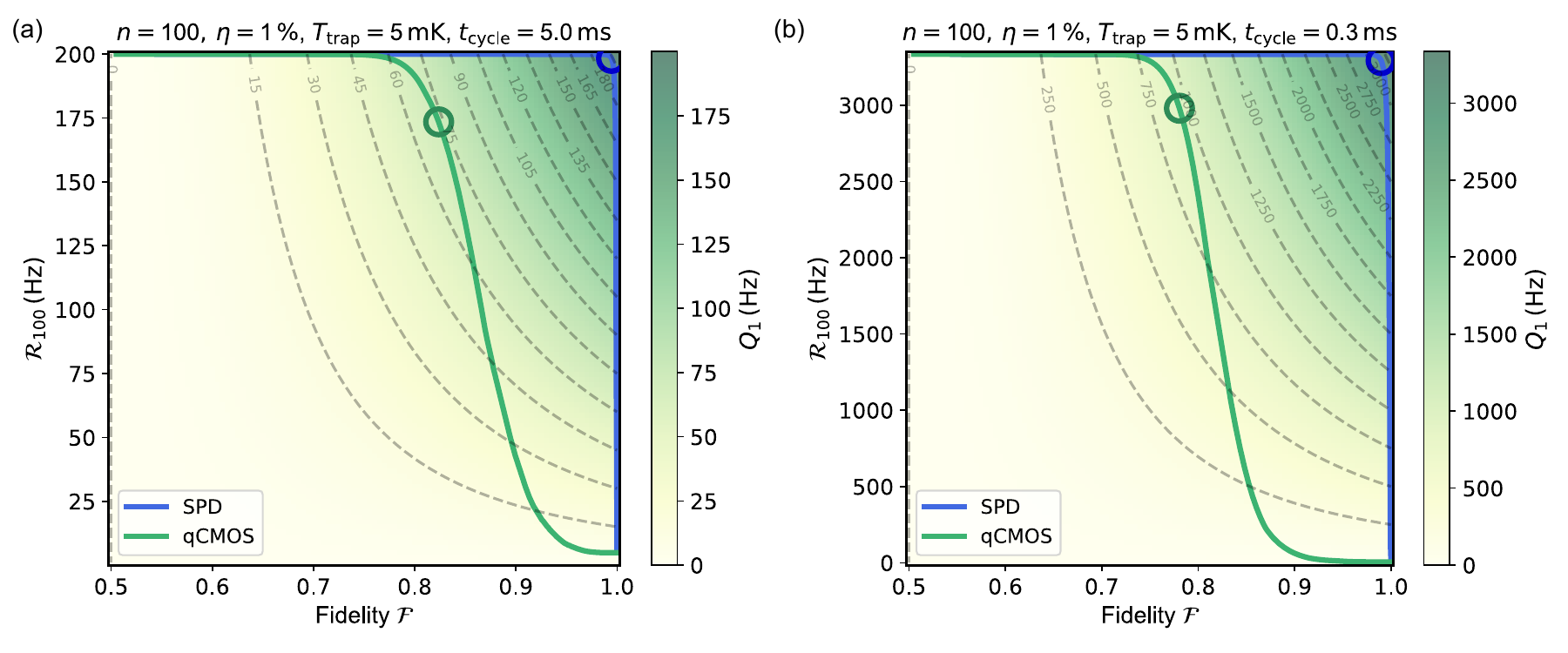}
\par\end{centering}
\caption{The trade-off characteristics between $\mathcal{R}_{100}$ and $\mathcal{F}$ for SPD and qCMOS with same atom number $n=100$ but different $t_\mathrm{cycle}$. The related readout device parameters are set to $(R_\mathrm{S},\,N_\mathrm{Q},\,\sigma_\mathrm{Q})=(500\,\mathrm{Hz},\,360,\,4)$, and related experimental parameters are set to $(t_\mathrm{dead},\,T_\mathrm{i},\,T_\mathrm{trap},\,\eta,\,R_\mathrm{sc})=(200\,\mathrm{ms},\,10\,\mu\mathrm{K},\,1\,\mathrm{mK},\,1\,\%,\,9.5\,\mathrm{MHz})$. The background color and the color bar indicate $Q_1$ at each point, and the contour lines are also plotted for reference. Points corresponding to optimal $Q_1$ on those curves are indicated by hollow circles. (a) $t_\mathrm{cycle}=5\,\mathrm{ms}$. By SPD, we have the optimal $Q_1=194.0\,\mathrm{Hz}$ for $\mathcal{R}_{100}=198.4\,\mathrm{Hz}$ and $\mathcal{F}=99.45\,\%$. While by qCMOS, we have the optimal $Q_1=72.7\,\mathrm{Hz}$ for $\mathcal{R}_{100}=173.5\,\mathrm{Hz}$ and $\mathcal{F}=82.37\,\%$. (b) $t_\mathrm{cycle}=0.3\,\mathrm{ms}$. By SPD, we have the optimal $Q_1=3158.0\,\mathrm{Hz}$ for $\mathcal{R}_{100}=3292.3\,\mathrm{Hz}$ and $\mathcal{F}=98.97\,\%$. While by qCMOS, we have the optimal $Q_1=938.1\,\mathrm{Hz}$ for $\mathcal{R}_{100}=2979.6\,\mathrm{Hz}$ and $\mathcal{F}=78.06\,\%$.}
\label{Fig7}
\end{figure*}

We further investigate the dependence of optimal achievable $Q_1$ on several parameters {[}see Fig.$\,$\ref{Fig6}{]} as a reference for parameter optimization. The curves in Fig.$\,$\ref{Fig6}(a) is calculated under $T_\mathrm{trap}=1\,\mathrm{mK}$, and curves in Fig.$\,$\ref{Fig6}(b) is calculated under $\eta=0.3\%$. These two figures exhibit similar trends yet display distinct differences. The approximations and analyses presented below can help elucidate this phenomenon and understand experimental parameter optimization.

If we ignore the time-dependent noise of the detector, $R_\mathrm{sc}$ and $\tau$ lose their independence, and $N_\mathrm{sc}=R_\mathrm{sc}\tau$ becomes the true hidden variable governing the tradeoff between $\mathcal{F}$ and $\mathcal{R}$. Under fixed device noise parameters, $\eta N_\mathrm{sc}$ uniquely determines $\mathcal{F}$. Furthermore, as analyzed in Sec.III.A, $\xi=T_\mathrm{a}/T_\mathrm{trap}=(T_\mathrm{i}+2N_\mathrm{sc} \Delta T_\mathrm{ph})/T_\mathrm{trap}$ uniquely determines the retention probability $P_\mathrm{ret}$ after once readout, which in turn dictates $\mathcal{R}$. If we further neglect the initial atomic temperature $T_\mathrm{i}$, this independent variable reduces to $2N_\mathrm{sc} \Delta T_\mathrm{ph}/T_\mathrm{trap}$. In this regime, $\eta$ and $T_\mathrm{trap}$ attain equivalent status in optimizing the $\mathcal{F}-\mathcal{R}$ tradeoff. Specifically, scaling $\eta$ or $T_\mathrm{trap}$ individually by the same factor will yield identical tradeoff curves between $\mathcal{F}$ and $\mathcal{R}$. Time-dependent noise diminishes the effectiveness of enhancing $T_\mathrm{trap}$, as requiring a longer $\tau$ to rescale the hidden variable, thereby accumulating higher noise. In contrast, enhancing $\eta$ avoids this issue. While the presence of the $T_\mathrm{i}$ diminishes the effectiveness of enhancing $\eta$, as enhancing $\eta$ does not improve the loss probability induced by $T_\mathrm{i}$. Owing that the noise of qCMOS is time-independent, only the effectiveness of enhancing $\eta$ is suppressed, which can be seen by comparing the required multiplication factor for $\eta$ and $T_\mathrm{trap}$ to reach the inflection point starting from $(\eta,\,T_\mathrm{trap})=(0.3\,\%,\,1\,\mathrm{mK})$. For SPD, these two effects both exist, so the curves behave more complexly. Considering that the difficulty of increasing $\eta$ and $T_\mathrm{trap}$ is different in practice, these conclusions are only for reference.

\section{Discussion}

Comparing SPD and qCMOS, the key advantage of SPDs lies in their lower noise and individual electronic channels, potentially behaving with lower latency for mid-circuit measurements and feedforward operations. However, when scaling up the number of qubits, qCMOS has the advantage of lower costs. In terms of implementation, SPDs are more suitable for high-precision, individual-state measurement scenarios. qCMOS cameras are better suited for large-scale experiments where the goal is to simultaneously capture data from a large number of atoms, sacrificing some readout precision for the ability to handle more atoms. To quantitatively compare the two approaches, we consider the cases with 100 atoms, optimizing $Q_1$ for adaptive or non-adaptive strategies. For the adaptive strategy, we reset the tweezer array when one or more atoms are lost, with the parameters of Fig.~\ref{Fig5}(d). Then, we have the optimal $Q_1=194.0\,\mathrm{Hz}$ for $\mathcal{R}_{100}=198.4\,\mathrm{Hz}$ and $\mathcal{F}=99.45\,\%$ by SPD, and $Q_1=72.7\,\mathrm{Hz}$ for $\mathcal{R}_{100}=173.5\,\mathrm{Hz}$ and $\mathcal{F}=82.37\,\%$ by qCMOS {[}see Fig.$\,$\ref{Fig7}(a){]}. For the non-adaptive strategy, we fix the rounds of cycles of $n$, and discard the results for the cycles in which one or more atoms are lost. By optimizing $n$ and readout duration, we have optimal $Q_1=188.8\,\mathrm{Hz}$ for $n=4149$, $\mathcal{R}_{100}=196.2\,\mathrm{Hz}$ and $\mathcal{F}=99.05\,\%$ by SPD, and $Q_1=62.8\,\mathrm{Hz}$ for $n=444$, $\mathcal{R}_{100}=168.7\,\mathrm{Hz}$ and $\mathcal{F}=80.49\,\%$ by qCMOS. Although the cases with SPD always outperform qCMOS, the qCIR by qCMOS is way higher than $1/t_{\mathrm{dead}}$, indicating a significant improvement in efficiency of the neutral atom processors in practical comparison with conventional schemes. Further improvement of the performance is possible by incorporating re-arrangement of the tweezer array by supplying the lost atoms from an atom reservoir, given that the transportation of a few atoms can be accomplished within $1\,\mathrm{ms}$ if the latency in data processing and electronic wires are suppressed~\citep{lam2021,bluvstein2024,lin2024}.

As indicated by Fig.~\ref{Fig5}(d), an $Q_1$ that is close to the ultimate limit imposed by $t_{\mathrm{cycle}}$, thus demands optimizing the experimental sequences in the practical experiments. First, the initialization process of a single atom in tweezer could be accelerated by phase-stable polarization gradient cooling or electromagnetically-induced-transparency (EIT) cooling, which can be accomplished in $100\,\mathrm{\mu s}$~\citep{feng2020,clements2024}. Second, the quantum gate sequence can be implemented within a duration of around $100\,\mathrm{\mu s}$ for a circuit depth of 100, given that the two-qubit gates and single-qubit gates could be implemented within a microsecond~\citep{levine2019,evered2023,bluvstein2024,radnaev2025}. Typical detection duration is around $50\,\mathrm{\mu s}$. Thus, a potentially achievable $t_\mathrm{cycle}$ could be suppressed to $300\,\mathrm{\mu s}$, experimentally feasible parameter set to $(\eta,\,T_{\mathrm{trap}})=(1\%,\,5\,\mathrm{mK})$ give rise to a highest achievable $Q_1=3158.0\,\mathrm{Hz}$ with $\mathcal{F}=98.97\,\%$ and $\mathcal{R}_{100}=3292.3\,\mathrm{Hz}$ by SPD, and $Q_1=938.1\,\mathrm{Hz}$ with $\mathcal{F}=78.06\,\%$ and $\mathcal{R}_{100}=2979.6\,\mathrm{Hz}$ by qCMOS {[}see Fig.$\,$\ref{Fig7}(b){]}.

It is worth noting that all analyses in this work are based on the assumption that only the photon recoil effect contributes to the heating of the atom in the trap. In practice, the increase of atom temperature is also determined by the detuning and power of the readout laser, as well as the alignment of the readout beam propagation direction, polarization, and the magnetic field. Although $\Delta T$ might vary in different experimental configurations, the relationship and trend of the readout strategies should be the same. In particular, the heating is determined by $N_\mathrm{sc}\Delta T$, and only the combined factor $(T_{\mathrm{trap}}-T_{\mathrm{i}})\eta/\Delta T$ contribute the performances of the system. For a higher $\Delta T$ than the simple recoil contribution, the conclusions and results are the same by scaling up the $\eta$ or $T_{\mathrm{trap}}-T_{\mathrm{i}}$. Or, a lower heating rate means a better performance equivalent to the improvement in $\eta$ or $T_{\mathrm{trap}}-T_{\mathrm{i}}$. In addition, the anti-trapped excited states resulting from the tweezer introduce another heating mechanism, not only leading to significantly higher $\Delta T$ than in the recoil effect but also linking $\Delta T$ to other experimental parameters, making it a variable instead of a constant~\citep{martinez-dorantes2018}. This effect severely complicates the analysis of the model, yet it can be addressed using techniques such as alternate switching of the tweezer and readout laser~\citep{nikolov2023,kwon2017,radnaev2025,bluvstein2024}, setting a far-detuned readout laser~\citep{martinez-dorantes2017,martinez-dorantes2018}, or utilizing magic trap wavelength~\citep{chow2023}. Therefore, we will not discuss the optimization of the readout strategy under conditions with this heating mechanism in detail here.

\section{Conclusion}

In conclusion, this work provides a comprehensive and quantitative framework for understanding and optimizing the trade-off during readout process in neutral atom quantum systems. By integrating physical models with quantum information metrics such as qCIR and normalized quantum Fisher information, we have developed practical guidelines applicable across a wide range of quantum computing and sensing platforms. Although the non-destructive readout methods have significant potential, they must be carefully balanced between factors like detection noise and readout duration to avoid compromising the overall performance. Ultimately, our study reveals the critical interplay between fidelity with retention probability, highlighting the potential for repeated readout without significant atomic loss. As neutral atom quantum computing continues to advance, these insights will be crucial for the development of scalable, high-performance quantum systems, facilitating the implementation of sophisticated quantum algorithms and improved sensing techniques.

\vbox{}

\section*{Acknowledgements}
\begin{acknowledgments}
This work was funded by the National Key R\&D Program (Grant No.~2021YFA1402004), the National Natural Science Foundation of China (Grants No. U21A20433, U21A6006, 92265210, and 92265108), and Innovation Program for Quantum Science and Technology (Grant No.~2021ZD0300203). This work was also supported by the Fundamental Research Funds for the Central Universities and USTC Research Funds of the Double First-Class Initiative. The numerical calculations in this paper have been done on the supercomputing system in the Supercomputing Center of University of Science and Technology of China. This work was partially carried out at the USTC Center for Micro and Nanoscale Research and Fabrication.
\end{acknowledgments}

\appendix
\section{\uppercase{signal distribution for qCMOS}}

For qCMOS camera, we can use the characteristic functions of the $\ket{0}$ state signal and atomic fluorescence signal to solve for the $\ket{1}$ state signal distribution. The characteristic functions of the $\ket{0}$ state signal and atomic fluorescence signal are as follows:
\begin{align}
\widetilde{M}_\mathrm{\ket{0},QC}(\omega)&=\int_{-\infty}^{+\infty}e^{in\omega}g_\mathrm{\ket{0},QC}(n)dn=e^{i\omega N_\mathrm{QC}-\frac{\sigma_\mathrm{QC}^{2}\omega^{2}}{2}},\label{eq:A1}\\
\widetilde{M}_\mathrm{a}(\omega)&=\sum_{n=0}^{+\infty}e^{in\omega}g_\mathrm{a}(n)dn=e^{\eta N_\mathrm{sc}\left( e^{i\omega}-1 \right)}.\label{eq:A2}
\end{align}
The $\ket{1}$ state signal is the sum of the $\ket{0}$ state signal and the atomic fluorescence, which are two independent random variables. Therefore, its characteristic function can be calculated as follows:
\begin{align}
\widetilde{M}_\mathrm{\ket{1},QC}(\omega)&=\widetilde{M}_\mathrm{\ket{0},QC}(\omega)\cdot\widetilde{M}_\mathrm{a}(\omega)=e^{i\omega N_\mathrm{QC}-\frac{\sigma_\mathrm{QC}^{2}\omega^{2}}{2}+\eta N_\mathrm{sc}\left( e^{i\omega}-1 \right)}\notag\\
&=e^{i(N_\mathrm{QC}\omega+\eta N_\mathrm{sc}sin\omega)-(\frac{\sigma_\mathrm{QC}^{2}\omega^{2}}{2}+2\eta N_\mathrm{sc}sin^{2}\frac{\omega}{2})}
,\label{eq:A3}
\end{align}
where the term $e^{-\sigma_\mathrm{QC}^{2}\omega^{2}/2}$ acts as an exponential decay factor, imposing a strong constraint on the range of values for $\omega$. We can approximate that Eq.~\ref{eq:A3} only counts when $\omega\in\left[ -2/\sigma_\mathrm{QC},\,2/\sigma_\mathrm{QC} \right]=\left[ -1/2,\,1/2 \right]$. Within this range, $\omega$ is small enough that we can approximate $sin\omega\approx\omega$. Then, Eq.~\ref{eq:A3} can be simplified as:
\begin{align}
\widetilde{M}_\mathrm{\ket{1},QC}(\omega)=e^{i(N_\mathrm{QC}+\eta N_\mathrm{sc})\omega-\frac{\sigma_\mathrm{QC}^{2}+\eta N_\mathrm{sc}}{2}\omega^{2}}
,\label{eq:A4}
\end{align}
indicating that the $\ket{1}$ state signal can be approximated as a Gaussian distribution with a mean of $(N_\mathrm{QC}+\eta N_\mathrm{sc})$ and a variance of $(\sigma_\mathrm{QC}^{2}+\eta N_\mathrm{sc})$:
\begin{align}
g_\mathrm{\ket{1},QC}(n)=\frac{1}{\sqrt{2\pi(\sigma_\mathrm{QC}^2+\eta N_\mathrm{sc})}} e^{-\frac{(n-N_\mathrm{QC}-\eta N_\mathrm{sc})^2}{2(\sigma_\mathrm{QC}^2+\eta N_\mathrm{sc})}} .\label{eq:A5}
\end{align}
The same as $g_\mathrm{\ket{0},QC}(n)$, we will discretize $g_\mathrm{\ket{1},QC}(n)$ by integrating over the integer intervals of this distribution in actual subsequent calculations.

\section{\uppercase{quantum Fisher information}}

Depending on a specific estimation task, the Fisher information $I$ describes the accuracy of this estimation by once sampling. $I$ has the same function of evaluating the information amount obtained by once sampling as $\mathcal{F}$, but it is more user-friendly in quantitative calculations, and its numerical significance is more intuitive and clear. Given a task of estimating a parameter $\theta$ by measuring the population $P_\mathrm{\ket{0}}(\theta)$ and $P_\mathrm{\ket{1}}(\theta)$, $I$ is defined by:

\begin{align}
I=\frac{1}{P_\mathrm{\ket{0}}(\theta)} \left[ \frac{\partial{P_\mathrm{\ket{0}}(\theta)}}{\partial{\theta}} \right] ^2 + \frac{1}{P_\mathrm{\ket{1}}(\theta)} \left[ \frac{\partial{P_\mathrm{\ket{1}}(\theta)}}{\partial{\theta}} \right] ^2.\label{eq:B1}
\end{align}

Given the constrain of $P_\mathrm{\ket{0}}+P_\mathrm{\ket{1}}=1$, we can simplify the equation:

\begin{align}
I=\frac{1}{P_\mathrm{\ket{1}}(\theta)(1-P_\mathrm{\ket{1}}(\theta))} \left[ \frac{\partial{P_\mathrm{\ket{1}}(\theta)}}{\partial{\theta}} \right] ^2.\label{eq:B2}
\end{align}

To focus on the dependence of $I$ on $\mathcal{F}$, we assume a specific task of estimating the Rabi frequency shift by directly measuring the population of a cosine-shaped Rabi curve at $\theta=\pi/2$. All subsequent calculations of $I$ will be based on this particular task. Then, we can further quantify the influence of $\mathcal{F}$ on $I$.

Considering the influence of $\mathcal{F}$, an atom with a probability of $P_\mathrm{\ket{1}}$ in $\ket{1}$ will be detected as in $\ket{1}$ with a probability of:

\begin{align}
P_\mathrm{\ket{1}}^*=\mathcal{F}P_\mathrm{\ket{1}}+(1-\mathcal{F})(1-P_\mathrm{\ket{1}}).\label{eq:B3}
\end{align}

Hence, the detected Rabi curve will be flattened from a 0-to-1 oscillation to a $(1-\mathcal{F})$-to-$\mathcal{F}$ oscillation. Taking Eq.~\ref{eq:B3} and $\theta=\pi/2$ into Eq.~\ref{eq:B2}, we can calculate the Fisher information $I$ of once sampling with fidelity $\mathcal{F}$ in this estimation task:

\begin{align}
I=\frac{1}{0.5*0.5}*(\frac{2\mathcal{F}-1}{2})^2=(2\mathcal{F}-1)^2,\label{eq:B4}
\end{align}

and the normalized quantum Fisher information $Q$:

\begin{align}
Q=\mathcal{R}I=\mathcal{R}(2\mathcal{F}-1)^2.\label{eq:B5}
\end{align}

Although the above derivation relies on a specific estimation task, the derived expression $I=(2\mathcal{F}-1)^2$ possesses deeper physical significance. Next, we will derive the same expression through another experimental-oriented analytical approach.

In experiments, we need to perform multiple executions of quantum circuits and readouts to obtain accurate results. The accuracy can be defined by the inverse of variance $D^{-1}$. 
Assuming the concerned information is encoded in the population of the $\ket{1}$ state $P_{\ket{1}}$ after circuit execution, for a perfect readout ($\mathcal{F}=1$), the inverse variance obtained from N times of readout would be:

\begin{align}
D^{-1}(P_{\ket{1}})=\frac{N}{P_{\ket{1}}(1-P_{\ket{1}})}.\label{eq:B6}
\end{align}

For an imperfect readout ($\mathcal{F}\textless1$) where the detected probability becomes $P_{\ket{1}}^*$. The imperfect readout compresses the population range from $\left[0,1\right]$ to $\left[1-\mathcal{F},\mathcal{F}\right]$. This error can be characterized a priori and corrected by scaling the experimental readout results. After correction, the inverse variance would correspondingly scale to:

\begin{align}
D^{-1}(P_{\ket{1}}^*)=\frac{N\left[\mathcal{F}-(1-\mathcal{F})\right]^2}{P_{\ket{1}}^*(1-P_{\ket{1}}^*)}=\frac{N(2\mathcal{F}-1)^2}{P_{\ket{1}}^*(1-P_{\ket{1}}^*)}.\label{eq:B7}
\end{align}

By setting the $D^{-1}$ of a perfect readout as a reference, we can evaluate the impact of readout fidelity $\mathcal{F}$ on information acquisition through the attenuation ratio $\kappa$ of $D^{-1}$ between imperfect and perfect readouts:

\begin{align}
\kappa(\mathcal{F})=&\frac{D^{-1}(P_{\ket{1}}^*)}{D^{-1}(P_{\ket{1}})}
=(2\mathcal{F}-1)^2\frac{P_{\ket{1}}(1-P_{\ket{1}})}{P_{\ket{1}}^*(1-P_{\ket{1}}^*)}\notag\\
=&\frac{(2\mathcal{F}-1)^2}{\frac{\mathcal{F}(1-\mathcal{F})}{P_{\ket{1}}(1-P_{\ket{1}})}+(2\mathcal{F}-1)^2}.\label{eq:B8}
\end{align}

When $P_{\ket{1}}=0.5$, we can derive $\kappa=(2\mathcal{F}-1)^2$, which matches the Fisher information expression given in Eq.~\ref{eq:B4}. However, the actual distribution of $P_{\ket{1}}$ after circuit execution may influence the form of $\kappa$. To establish a fair evaluation, we calculate the expected value of $\kappa$ under uniformly distributed final states on the Bloch sphere:

\begin{align}
\kappa(\mathcal{F})=\int_{0}^{1}\frac{(2\mathcal{F}-1)^2}{\frac{\mathcal{F}(1-\mathcal{F})}{P_{\ket{1}}(1-P_{\ket{1}})}+(2\mathcal{F}-1)^2}dP_{\ket{1}}.\label{eq:B9}
\end{align}

\begin{figure}[t]
\begin{centering}
\includegraphics[width=1\linewidth]{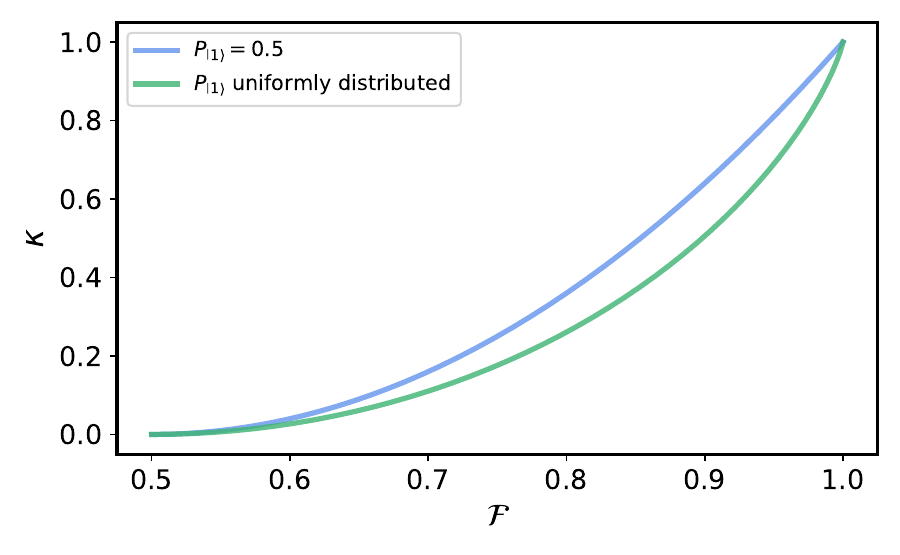}
\par\end{centering}
\caption{Calculated dependence of the information attenuation ratio $\kappa$ on readout fidelity $\mathcal{F}$.}
\label{FigB}
\end{figure}

As this expression lacks an analytical solution, we numerically investigate its properties and compare with $(2\mathcal{F}-1)^2$. The results is shown in Fig.$\,$\ref{FigB}, where the blue curve corresponds to the case of $P_{\ket{1}}=0.5$ (i.e., $\kappa=I=(2\mathcal{F}-1)^2$), while the green curve represents the expectation value of $\kappa$ when $P_{\ket{1}}$ is uniformly distributed over the Bloch sphere.

These two curves exhibit close numerical agreement and follow similar trends, with even the maximum deviation showing only approximately a $20\,\%$ relative difference. Therefore, we propose that $I=(2\mathcal{F}-1)^2$ serves as a concise and effective metric for evaluating how readout fidelity impacts single-shot information acquisition. While incorporating the distribution of $P_{\ket{1}}$ could improve accuracy, it does not undermine the validity of the conclusions or numerical results presented in the main text.

\section{\uppercase{experimental parameters}}

For clarity and reliability, this section presents the typical ranges of experimental parameters used in our calculations under current experimental techniques, along with relevant works. To facilitate experimental implementation, we further specify estimation protocols for each parameter in practical settings.

\begin{enumerate}
    \item Photon scattering rate $R_\mathrm{sc}$. In principle, $R_\mathrm{sc}$ can vary from 0 to $\Gamma/2$, where $\Gamma$ is the transition linewidth for readout. Yet, to avoid inducing excessive stray light, the readout laser intensity is typically set to around saturation intensity~\citep{martinez-dorantes2017,nikolov2023,kwon2017,chow2023}, corresponding to a scattering rate of $\Gamma/4$. Hence, we choose a scattering rate $R_\mathrm{sc}=\Gamma/4\approx9.5\,\mathrm{MHz}$ for $^{87}\mathrm{Rb}$.
    \item Collection efficiency $\eta$. For cavity-free neutral atom systems, $\eta$ is typically around $1\,\%$~\citep{gibbons2011,chow2023,shea2020}. By employing high-NA lenses and techniques such as multi-lens collection, $\eta$ can achieve nearly $10\,\%$~\citep{graham2022,radnaev2025}.
    \item Initial atomic temperature $T_\mathrm{i}$. Utilizing sub-Doppler cooling techniques such as polarization gradient cooling, grey molasses, or EIT cooling, $T_\mathrm{i}$ can typically be suppressed below $100\,\mu\mathrm{K}$~\citep{tuchendler2008,fuhrmanek2011,chow2023,bluvstein2024,radnaev2025,Brown2019,nikolov2023,kwon2017,chow2023}.
    \item Trap depth $T_\mathrm{trap}$. To achieve non-destructive readout, a high trap depth is usually needed, with typical values ranging from one or several millikelvens~\citep{gibbons2011,martinez-dorantes2017,chow2023,shea2020,bluvstein2024,radnaev2025} to over $10\,\mathrm{mK}$~\citep{nikolov2023,kwon2017}.
    \item Dark count rate for SPD $R_\mathrm{D,SPD}$. $R_\mathrm{D,SPD}$ varies significantly among individual devices, with typical values around hundreds of hertz~\citep{gibbons2011,chow2023,shea2020,fuhrmanek2011}. However, SPDs with dark count rates as low as tens of hertz are also available~\citep{bochmann2010,hu2024siteselective}.
    \item Variance of the readout noise for qCMOS $\sigma^2_\mathrm{QC}$. While research utilizing qCMOS camera remains limited to date, the data employed in this paper are estimated from data and figures in Reference~\citep{manetsch2024,bluvstein2024}.
\end{enumerate}

$R_\mathrm{sc}$ and $\eta$ can be simultaneously estimated by measuring the dependence of collected photon rate $R_\mathrm{collect}$ on the intensity of a resonant readout laser $I_\mathrm{ro}$, and fitting the data to the expression $R_\mathrm{collect}=\eta\frac{\Gamma}{2}\frac{kI_\mathrm{ro}}{kI_\mathrm{ro}+I_\mathrm{sat}}$, where $k$ is a coefficient for laser misalignment, $I_\mathrm{sat}$ is the saturation intensity~\citep{Shea2018Fast}. $T_\mathrm{i}$ can be estimated by the "release-and-recapture" protocol~\citep{tuchendler2008}. $T_\mathrm{trap}$ can be estimated by measuring trap frequencies~\citep{friebel1998} using parametric excitation~\citep{wu2006}. The noise characteristics of detectors can be obtained by measuring the signal without loaded atoms.

Besides, experimental researchers can directly measure the trade-off relationship between $P_\mathrm{ret}$ and $\mathcal{F}$ by scanning the readout stage duration $\tau$, and subsequently apply the methodology outlined in Sec. III. C to analyze and identify the experimental parameters that maximize the system's information acquisition rate.

\end{document}